\documentclass [prb,reprint,notitlepage,superscriptaddress,floatfix] {revtex4-1}
\usepackage{amsmath}
\usepackage{graphicx}
\usepackage{lmodern}
\usepackage{amsmath}
\usepackage{color}

\usepackage{amssymb}
\newcommand{\be} {\begin{align}}
\newcommand{\ee} {\end{align}}
\renewcommand{\(}{\left(}
\renewcommand{\)}{\right)}
\renewcommand{\[}{\left[}
\renewcommand{\]}{\right]}

\DeclareMathOperator{\Tr}{Tr}

\begin{document}

\title {Interplay between {short-range correlated} disorder and Coulomb interaction in nodal-line semimetals}
\author{Yuxuan Wang}
\affiliation{Department of Physics and Institute for Condensed Matter Theory, University of Illinois at Urbana-Champaign, Urbana, Illinois 61801, USA }
\author{Rahul M. Nandkishore}
\affiliation{Department of Physics and Center for Theory of Quantum Matter, University of Colorado, Boulder, Colorado 80309, USA}
\date{\today}
\begin{abstract}
In nodal-line semimetals, Coulomb interactions and {short-range correlated} disorder are both marginal perturbations to the clean non-interacting Hamiltonian. We analyze their interplay using a weak-coupling renormalization group approach.  In the clean case, the Coulomb interaction has been found to be marginally irrelevant, leading to Fermi liquid behavior. We extend the analysis to incorporate the effects of disorder. The nodal line structure gives rise to kinematical constraints similar to that for a two-dimensional Fermi surface, which plays a crucial role in the one-loop renormalization of the disorder couplings. For a two-fold degenerate nodal loop (Weyl loop), we show that disorder flows to strong coupling along a unique fixed trajectory in the space of symmetry inequivalent disorder couplings. Along this fixed trajectory, all symmetry inequivalent disorder strengths become equal. For a four-fold degenerate nodal loop (Dirac loop), disorder also flows to strong coupling, however the strengths of symmetry inequivalent disorder couplings remain different.  We show that feedback from disorder reverses the sign of the beta function for the Coulomb interaction, causing the Coulomb interaction to flow to strong coupling as well. However, the Coulomb interaction flows to strong coupling asymptotically more slowly than disorder. Extrapolating our results to strong coupling, we conjecture that at low energies nodal line semimetals should be described by a noninteracting nonlinear sigma model. We discuss the relation of our results with possible many-body localization at zero temperatures in such materials.
\end{abstract}
\maketitle

\section{Introduction}
Topological semimetals~\cite{vafek-vishwanath,burkov16}  host a gapless fermionic spectrum with vanishing density of states at a certain filling. 
Unlike topological insulators, the presence of gapless excitations in topological semimetals  offers a unique platform for the interplay of symmetry, topology, interaction and disorder effects (for example, see Refs.\ \onlinecite{weyl-interaction1, weyl-interaction2,Fradkin,Fradkin2, NandkishoreHuseSondhi, subir-qed3, SyzranovARCMP, sri-qed3}). While early studies focused on graphene~\cite{aleiner-efetov-06,foster-aleiner-08, Graphenereview1, Graphenereview2}, recently the prediction and discovery of Weyl and Dirac semimetals~\cite{weyl1,weyl2, Bernevig1, dirac1,dirac2} has generated great interest in properties of semimetals in three dimensions. 

Nodal-line semimetals \cite{Carteretal, ChenLuKee, Schafferetal, KimWiederKaneRappe, Mullenetal, aletFu, aletHu, okamoto, weng-16, Cava,NeupertHasan,Takenaka} are a new class of three dimensional topological semimetal, distinct from Weyl and Dirac point semimetals. Nodal line semimetals have Fermi surfaces of co-dimension two, and contain {\it lines} of Dirac points. As such, they have properties intermediate between graphene (which also has Fermi surface codimension two) and Weyl semimetals (which also exist in three dimensions). The nodal line can either involve a crossing of two bands (Weyl) or a crossing of two doubly degenerate bands (Dirac), and may either stretch across the Brillouin zone, or close inside the Brillouin zone forming a ``loop" of Dirac nodes. We will focus on the case of nodal loops, which are also associated with``drumhead" surface states ~\cite{BurkovHookBalents,YZ1, YZ2, YZ3,volovik2}. Nodal-loop semimetals 
have been experimentally realized in Ca$_3$P$_2$~\cite{Cava}, CaAgP and CaAgAs~\cite{Takenaka, Loop1}, with other candidate materials including TlTaSe$_2$~\cite{NeupertHasan} and CaP$_3$ family of materials~\cite{weng-16}, which constitute an exciting new chapter in the study of topological semimetals. 

Theoretical studies of nodal-loop semimetals are still in their infancy. While many studies have focused on the possibilities that such materials may host exotic correlated or superconducting states \cite{volovik2, WeylLoopSC, Roy, SurNandkishore, WangNandkishore, Jianpeng, HassanWangFuture} 
the zeroth order question of understanding the effect of Coulomb interactions and disorder in this setting remains open. For {\it clean} systems, the long-range Coulomb interaction was found to be marginally irrelevant~\cite{HuhMoonKim}, giving rise to Fermi liquid behavior at low-energies. However, realistic samples will inevitably contain disorder, and  by simple power counting, in nodal-line semimetals {short-range correlated} disorder and Coulomb interactions both have the same scaling dimension (marginal) at tree level. {In real materials, short-range correlated disorder can be introduced by screened Coulomb impurities~\cite{skinner}, or by lattice defects.}
As such, in a generic scenario the long-range Coulomb interaction and disorder need to be treated on an equal footing~\cite{NandkishorePara}. A controlled treatment of the interplay of Coulomb interactions and disorder in nodal loop semimetals is the primary focus of this work. 

The interplay between interaction and disorder has been the focus of enormous theoretical and experimental efforts, most notably the recent developments in many-body localization (MBL) (see Ref.\ \onlinecite{NandkishoreHuse} for a review). While the research in MBL mainly focuses on high or even infinite temperatures, the fate of disorder and interaction effects at zero temperature is also generally an interesting problem. In this paper we address the issue in the context of nodal-line semimetals using an weak-coupling renormalization group (RG) approach. At first sight, this problem seems quite similar to the analagous intensively studied problem in graphene~\cite{Stauber, SachdevYe, foster-aleiner-08}, where also both Coulomb interactions and disorder are marginal at tree level.  However, the important difference here is that instead of a discrete set of Dirac points, there exists a continuous angular degree of freedom for the low-energy fermions. This is in some sense akin to a two-dimensional Fermi surface~\cite{shankar}, but here with zero density of states. As we shall see, this angular degree of freedom plays an important role in distinguishing  different diagrams and  greatly simplifies the problem, reducing the important interactions to those in the forward-scattering and Cooper channels alone. Additionally, the dispersions of the Coulomb interaction are different in Weyl loop materials as compared to graphene: in two dimensions the interaction in momentum space takes a non-analytic $1/|q|$ form, while in three dimensions it takes the form $1/q^2$. As a consequence, and unlike the case of graphene~\cite{foster-aleiner-08}, it turns out that for a two-band model with a nodal loop (i.e., Weyl loop), {\it all} disorder types are marginally relevant and flow to strong coupling, and their ratios all tend to one. As we mentioned, in the clean case the Coulomb interaction strength flows to zero. However, we find that feedback from disorder reverses the sign of the Coulomb beta function and drives the Coulomb interaction also to strong coupling. Nevertheless, the ratio between the Coulomb interaction strength and the disorder strength flows to zero. It is thus reasonable to expect that the strong-coupling theory should be well described by an non-interacting nonlinear sigma model. This result is qualitatively similar to an earlier work by one of us on systems with quadratic band crossing points \cite{NandkishorePara}. However, for quadratic band crossings both interaction and disorder are {\it relevant} at tree level and to control the calculations one has to rely on an dimensional $\epsilon$ expansion. Here both disorder and Coulomb interactions are marginal, and no $\epsilon$ expansion is necessary. Additionally, the extended ``nodal loop" gives rise to kinematic constraints not present for a quadratic band crossing point, and thus not discussed in Ref.\ \onlinecite{NandkishorePara}. 
 
 The rest of the paper is organized as follows. In Sec.\ \ref{sec:2} we briefly recapitulate the essential results in the clean case with Coulomb interaction, and reproduce the results obtained in Ref.~\onlinecite{HuhMoonKim}. In Sec.\ \ref{sec:3} we present the results of RG analysis for different types of disorder in a non-interacting limit, and show that disorder flows to strong coupling, while the ratio of different disorder strengths flows to one. Note that we consider only short range correlated disorder - long range correlated disorder has been discussed in e.g. \cite{SkinnerSyz}. In Sec.\ \ref{sec:4} we address the issue of the feedback effects between interaction and disorder, and show that disorder drives the Coulomb interaction, which would vanish under its own RG flow, to strong coupling instead, while remaining asymptotically weaker than the disorder. In Sec.\ \ref{sec:5}, we briefly discuss the results for Dirac loop materials in comparison with Weyl loop materials. Finally  in Sec.\ \ref{sec:6} we discuss the experimental and theoretical implications of our results and present our conclusions.

\section{Coulomb interaction in a clean system}
\label{sec:2}

In this section, we discuss the renormalization of the Coulomb interaction for a two-band nodal loop model. The analysis follows Ref.\ \onlinecite{HuhMoonKim}. 

For simplicity we begin by considering a two band Weyl loop system, the simplest band structure that exhibits a nodal loop. More complicated nodal loop structures, such as a Dirac loop, can be thought of multiple copies of Weyl loops. The single-particle Hamiltonian can be written as
\begin{align}
\mathcal{H}_0=\frac{k_x^2+k_y^2-k_F^2}{2m}\sigma^x+v_z k_z \sigma^{y},
\label{wloop}
\end{align}
The system has a time reversal symmetry $\mathcal{T}=K$ {($K$ is complex conjugate operator)}, with $\mathcal{T}^2=1$.
We define $k_F/m\equiv v_F$ for future use.
We decompose the four-fermion static Coulomb interaction into fermionic bilinears coupled to a static bosonic field $\phi$ (in some sense a photon), and the full action in the disorder-free case can be written as
\begin{align}
\!\!\!\mathcal{S}=\sum_{i}^n\int d\tau d^3x \[\psi^\dagger (\partial_\tau+\mathcal{H}_0 + ie\phi_i)\psi +\frac{c}{2}(\nabla \phi_i)^2\] 
\label{action}
\end{align}
where we have kept a replica index $i$, which will be important when we consider the disorder problem later.

The renormalization group is performed on cylindrical  momentum shells around $k_r=k_F$ (where the nodal line is), with $|k_r-k_F|\in (\Lambda e^{-d\ell}, \Lambda)$. This scheme is the same as that used in Ref.\ \onlinecite{HuhMoonKim}, while another scheme, where the momentum shell forms a circle at any $\theta$ around the nodal line has been employed in Ref.\ \onlinecite{SurNandkishore}. Of course, the final result should not depend on the specific RG scheme. We convert the action \eqref{action} to momentum and frequency space and count the classical scaling dimensions of various operators. One important subtlety~\cite{HuhMoonKim} is that, for the fermion sector the momentum rescales towards the nodal line, and during this process the momentum along the angular direction (the component parallel to the nodal ring) does not get rescaled, since during the process the circumference of the nodal ring does not change. As a result, for the fermion sector $[d^3{\bf k}] = 2$. For the boson sector all three momentum components rescale toward 0, and thus for the bosonic sector $[d^3{\bf q}]=3$ [we will use $\bf k$ ($\bf q$) to denote fermionic (bosonic) momentum]. For generality, we take the frequency scaling as $[d\omega]=z$, and at tree level $z=1$.
Taking these into account, at the tree level we have for the Fourier transformed fields
\begin{align}
[\psi_{{\bf k},{\omega}}]= -1-z, ~[\phi_{{\bf q},\Omega}]= -\frac{5}{2}-\frac{z}{2}
\end{align}
From the Yukawa coupling term
\begin{align}
\mathcal{S}_{\psi\phi}=ie\sum_{i}^n \int d\omega d\omega' d^3{\bf k} d^3{\bf q} ~( \phi_{q} \psi_k^\dagger \psi_{k-q} ),
\end{align}
we can read off that
\begin{align}
[e]=\frac{z-1}{2},
\end{align}
and the Coulomb interaction is marginal at tree level. Note that the integration measure cannot be written in terms of two fermionic momenta $d^3{\bf k}d^3{\bf k'}$, since that way the high energy momentum bosons with $|{\bf q}|\sim 2k_F$ will always be involved throughout the RG flow.

We start with the boson self-energy in the static limit, given by
\begin{align}
\Pi({\bf q},0)=-(ie)^2\int_{{\bf k},\Omega}\Tr\[G_0(k+\frac{q}{2},\Omega)G_0(k-\frac{q}2,\Omega)\],
\end{align}
where the Green's function is given by 
\begin{align}
G_0(k\pm \frac{q}{2})=\frac{i\Omega+\sigma^x \epsilon_{1\pm}+\sigma^y \epsilon_{2\pm}}{\Omega^2+E_{\pm}^2},
\end{align}
where $\epsilon_{1\pm}\equiv v_F(k_r\pm\frac{1}{2}q_r\cos\theta)$ ($\theta$ is the relative angle between ${\bf k}_r$ and ${\bf q}_{r}$), $\epsilon_{2\pm}\equiv v_z (k_z\pm q_z/2)$, and $E_{\pm}=\sqrt{\epsilon_{1\pm}^2+\epsilon_{2\pm}^2}$. Evaluating the trace over spin indices and integrating over $\Omega$, we obtain \begin{align}
\Pi({\bf q},0)=&e^2\int_k\frac{-\Omega^2+\epsilon_{1+}\epsilon_{1-}+\epsilon_{2+}\epsilon_{2-}}{(\Omega^2+E_+^2)(\Omega^2+E_-^2)}\nonumber\\
=&-e^2\int_k\frac{E_+E_--(\epsilon_{1+}\epsilon_{1-}+\epsilon_{2+}\epsilon_{2-})}{E_+E_-(E_++E_-)}.
\label{rpa}
\end{align}
This integral can be evaluated after we expand the integrand in $q$:
\begin{align}
E_+E_--&\epsilon_{1+}\epsilon_{1-}-\epsilon_{2+}\epsilon_{2-}\nonumber\\
&\approx\frac{1}{2}\frac{\epsilon_2^2}{\epsilon_1^2+\epsilon_2^2}v_F^2{{q}_r}^2\cos^2 \theta+\frac{1}{2}\frac{\epsilon_1^2}{\epsilon_1^2+\epsilon_2^2}v_z^2q_z^2,
\end{align}
where $\epsilon_1=v_F k_r$, and $\epsilon_2=v_z k_z$. We only expand the integrand up to $q^2$ order, although higher order terms are also generated. By dimension counting such terms are irrelevant, and are thus unimportant to us. For the same reason we will expand the fermionic self-energy to linear order in $\bf k$ below.
We see  in \eqref{rpa} that $E_+E_--\epsilon_{1+}\epsilon_{1-}-\epsilon_{2+}\epsilon_{2-}$ is small in $q$ and to leading order we can set in the denominator $E_+=E_-=\sqrt{\epsilon_1^2+\epsilon_2^2}$.

\begin{figure*}
\includegraphics[width=1.2\columnwidth]{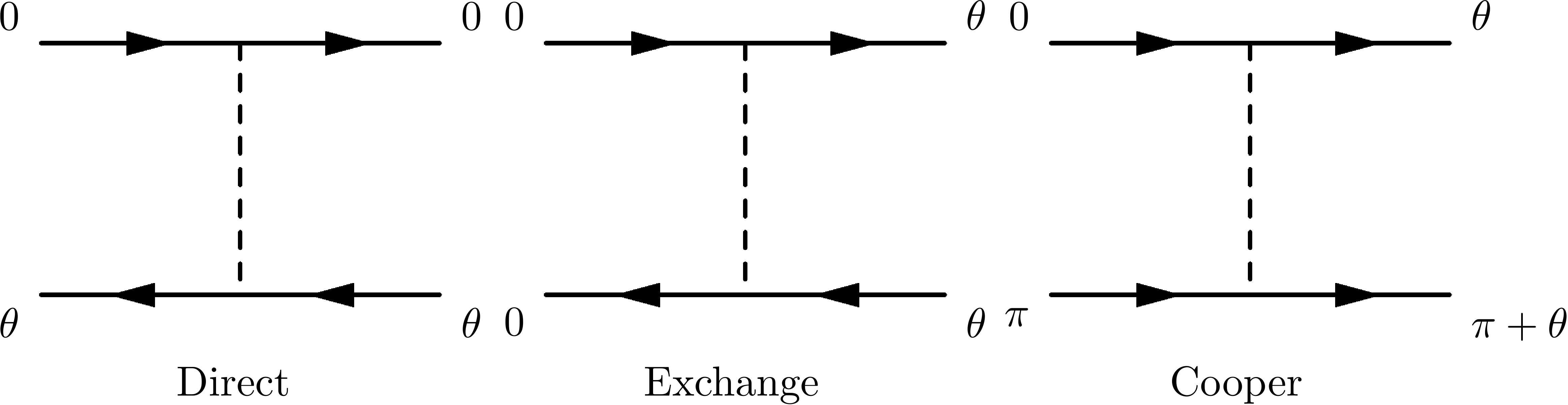}
\caption{The three channels of disorder coupling for low energy fermions, where the labels denote the angle along the Fermi ring. The placing of angle 0 is arbitrary.}
\label{fig1}
\end{figure*}

As we discussed, in the spirit of RG, the $k$-integral is done on the two cylindrical  momentum shells around $k_r=k_F$, with $|k_r-k_F|\in (\Lambda e^{-d\ell}, \Lambda)$, i.e., $\int_k=\Lambda d\ell k_F/(4\pi^2)\int_{-\infty}^{\infty}{dk_z} $.  Thus,
\begin{align}
\Pi({\bf q},0)=&-\frac{e^2\Lambda d\ell k_F}{2\pi^2v_z}\[\frac{1}{2}\int d\epsilon_2\frac{\epsilon_2^2}{4(v_F^2\Lambda^2+\epsilon_2^2)^{5/2}}v_F^2q_r^2\right.\nonumber\\
&~~~~~~~~~~~~~~~~~\left.+\int d\epsilon_2\frac{v_F^2\Lambda^2}{4(v_F^2\Lambda^2+\epsilon_2^2)^{5/2}}v_z^2q_z^2\] \nonumber\\
=&-\frac{e^2\Lambda d\ell k_F}{2\pi^2v_z}\(\frac{v_F^2 q_r^2}{12v_F^2\Lambda^2}  + \frac{v_z^2 q_z^2}{3v_F^2\Lambda^2} \) \nonumber\\
=& -\frac{e^2}{4\pi^2}\(q_r^2\frac{k_F}{6v_z}+q_z^2\frac{2m^2v_z}{3k_F}\)\frac{d\ell}{\Lambda}.
\nonumber\\
=& -{\bar \alpha}\[q_r^2\frac{1}{2\eta}+2\eta{q_z^2}\]{d\ell},
\label{110}
\end{align}
where in the first step, the $1/2$ in the first term comes from angular integration over $\int d\theta \cos^2 \theta$, and in the last step we have defined 
\begin{align}
{\bar \alpha}=\frac{e^2k_F}{12\pi^2v_F\Lambda(\ell)},~~\eta\equiv \frac{v_z}{v_F},
\label{def1}
\end{align} 
and 
\begin{align}
[\bar\alpha]=z.
\end{align}
Physically, $\bar\alpha$ is roughly the ratio of the Coulomb energy  $e^2 k_F$ and the cutoff energy $v_F\Lambda$.

The renormalized bosonic propagator is given by
\begin{align}
D({\bf q})=\frac{-e^2}{a q_r^2+q_z^2/a-\Pi(q_r,q_z)},
\end{align}
where we have included an anisotropy factor $a$ to allow for its renormalization. We emphasize also that that the anisotropy $a$ is {\it not} an artifact of the RG scheme with a cylindrical momentum shell, but rather a result of the global structure of the nodal loop, {for which the two directions perpendicular to the nodal loop are inequivalent}. Had we used a more symmetric ``torus" momentum shell RG scheme like the one used in Ref.\ \onlinecite{SurNandkishore}, the anisotropy would still be present. 

The polarization operator, being negative, contributes to a downturn renormalization (screening) of the Coulomb interaction, as well as the renormalization of $a$. We have so far only considered the bosonic propagator in the static limit. It was found in Ref.\ \onlinecite{HuhMoonKim} that the frequency dependence corresponds to the (weak) damping of the boson $\phi$. It does not contribute to the RG flow and is not of interest to us here.

Next let's look at the fermionic self-energy. It is clear that $\Sigma(\omega,{\bf k}_F+{\bf k})$ has no frequency dependence:
\begin{align}
\Sigma(\omega,{\bf k}_F+{\bf k})=\int_{\Omega,{\bf q}}\frac{i(\omega+\Omega)+\sigma^x\epsilon_1+\sigma^y\epsilon_2}{(\omega+\Omega)^2+E^2}\nonumber\\
\times\frac{-e^2}{a(q_x^2+q_y^2)+\frac{1}{a}q_z^2},
\end{align}
and integrating over $\Omega$ yields,
\begin{align}
\Sigma({\bf k})=&{-\frac{e^2}2}\int_{\bf q}\frac{v_F(k_x+q_x)\sigma^x+v_z(k_z+q_z)\sigma^y}{\sqrt{v_F^2(k_x+q_x)^2+v_z^2(k_z+q_z)^2}}\nonumber\\
&\times\frac{1}{a(q_x^2+q_y^2)+\frac{1}{a}q_z^2}.
\end{align}
Expanding for small $\bf k$ and integrating over $\bf q$, we have for the $k_x$ and $k_z$ dependence of $\Sigma$,
\begin{align}
\Sigma(k_F+k_x,0,0)=&-\frac{\alpha d\ell}{8\pi^2}\int_{z}\frac{a^2\eta^2 z^2dz}{(a^2\eta^2z^2+1)^{3/2}}\frac{1}{\sqrt{1+z^2}}\nonumber\\
\equiv & -\frac{\alpha d\ell}{8\pi^2}F_1(a\eta)\sigma_xv_F k_x,\nonumber\\
\Sigma(k_F, 0, k_z)=&-\frac{\alpha d\ell}{8\pi^2}\int_{z}\frac{dz}{(a^2\eta^2z^2+1)^{3/2}}\frac{1}{\sqrt{1+z^2}}\nonumber\\
\equiv & -\frac{\alpha d\ell}{8\pi^2}F_2(a\eta)\sigma_yv_z k_z
\label{13}
\end{align}
where  $z=q_r/q_z$ and
\begin{align}
\alpha\equiv e^2/v_F = 12\pi^2\bar \alpha \Lambda/k_F,
\end{align} and
\begin{align}
F_1(a\eta)&=\frac{2}{1-a^2\eta^2}E(1-a^2\eta^2)-\frac{2a^2\eta^2}{1-a^2\eta^2}K(1-a^2\eta^2),\nonumber\\
F_2(a\eta)&=\frac{2}{1-a^2\eta^2}E(1-a^2\eta^2)-\frac{2}{1-a^2\eta^2}K(1-a^2\eta^2),
\end{align}
where $K$ and $E$ are elliptic functions. From the self-energy the Fermi velocity acquires a scaling dimension $[v_F]=z-1+\alpha F_1/8\pi^2$. Enforcing the invariance of $v_F$, we obtain
\begin{align}
z=1-\frac{\alpha}{8\pi^2} F_1(a\eta).
\label{117}
\end{align}
On the other hand, this self-energy also renormalizes the ratio $\eta\equiv v_z/v_F$, which we address below.


Another important diagram for the renormalization of the Coulomb interaction is the vertex correction. However, as found in Ref.\ \onlinecite{HuhMoonKim}, the vertex correction vanishes at one-loop level.
With the results in Eqs.\ (\ref{110},~\ref{13}) the $\beta$-functions for the couplings and anisotropy parameters are
\begin{align}
\frac{d{\bar \alpha}}{d\ell}&={\bar \alpha}\[1-\frac{\alpha}{8\pi^2}F_1(a\eta)\]-\frac{\bar \alpha^2}{2}\(\frac{1}{2a\eta}+2a\eta\),\nonumber\\
\frac{da}{d\ell}&=\frac{a\bar\alpha}{2}\(\frac{1}{2a\eta}-2a\eta\)\nonumber\\
\frac{d\eta}{d\ell}&=\frac{\eta\alpha}{8\pi^2}\(F_2(a\eta)-F_1(a\eta)\).
\label{119}
\end{align}
Combining the last two lines,
\begin{align}
\frac{d(a\eta)}{d\ell}&=a\eta\[\frac{\bar\alpha}{2}\(\frac{1}{2a\eta}-2a\eta\)+\frac{\alpha}{8\pi^2}\(F_2(a\eta)-F_1(a\eta)\)\].
\end{align}
There are a few factor of 2 and sign differences compared with the result in Ref.\ \onlinecite{HuhMoonKim}, but only in the subleading terms $\sim \alpha\ll \bar \alpha$ in the brackets, which does not lead to any differences in the fixed points. The nontrivial fixed point is $\alpha=0, {\bar \alpha}=1, a\eta=1/2$.  Therefore, without disorder, the system flows to an noninteracting fixed point. On the other hand, from the renormalized propagator 
\begin{align}
D^{-1}\sim a\(1+\frac{{\bar \alpha}}{2a\eta} d\ell\)q_r^2+(1+{\bar \alpha}2a\eta d\ell) q_z^2,
\end{align} 
we see that the Coulomb interaction has an anomalous dimension of one, and at the IR limit, the dispersion of $\phi$ becomes linear, which is confirmed by a direct calculation in  the   large-$N$ limit~\cite{HuhMoonKim}.

\section{Renormalization group analysis for disorder}
\label{sec:3}

\begin{figure}
\includegraphics[width=0.75\columnwidth]{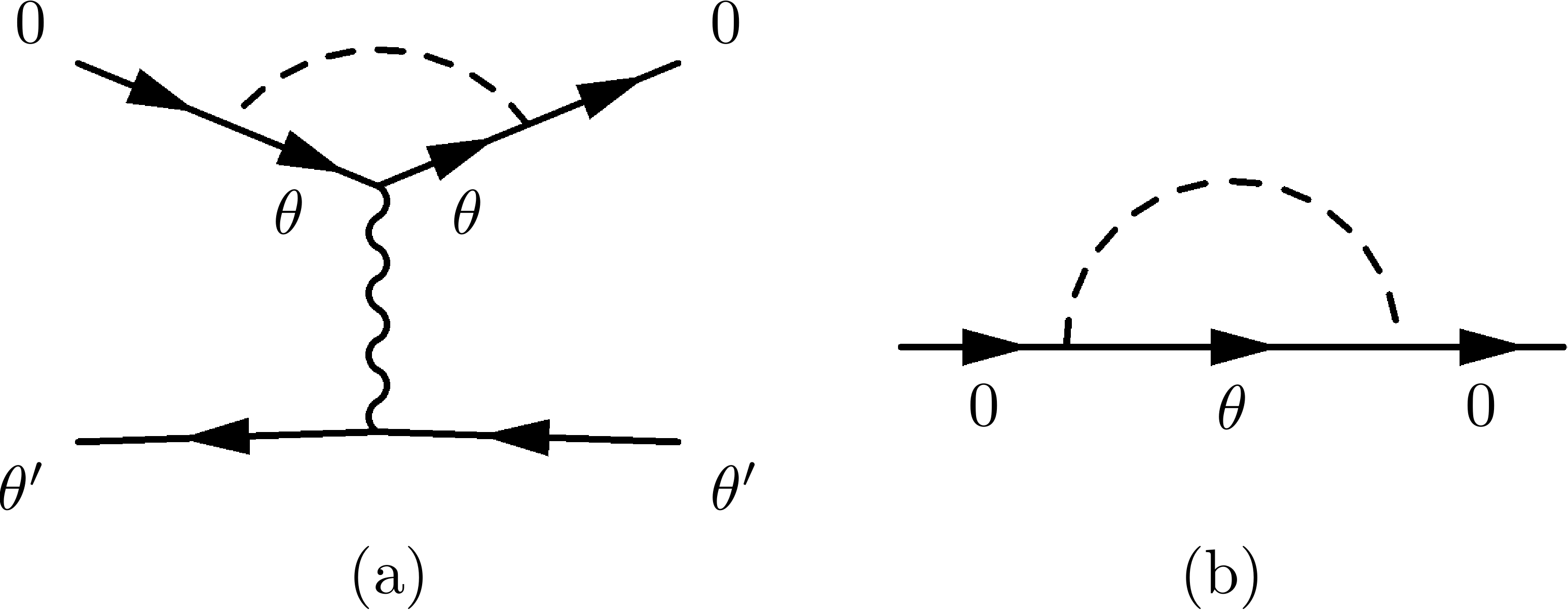}
\caption{Panel (a): The vertex correction to Coulomb interaction from disorder coupling in the exchange channel. Panel (b): The contribution to self-energy from disorder coupling in the exchange channel.}
\label{fig2}
\end{figure}

In this section, we consider disorder effects within the replica field theory formalism. For simplicity we temporarily neglect the effects of Coulomb interaction.
As we discussed in the Introduction, the angular degree of freedom of the nodal loop plays an important role. Similar to the case of a 2D Fermi surface, only the forward scattering (including direct and exchange channels) and back scattering (Cooper channel) can survive the RG process---all other channels are kinematically suppressed. We show these three channels in Fig.\ \ref{fig1}. Particularly we note that since along a fermion line the replica index is conserved, the direct channel and the exchange channel can be distinguished by keeping track of the replica indices, and thus can be separately treated.

The renormalization of the Coulomb interaction by disorder proceeds (at one loop level) only through the vertex correction (VC) diagram with the disorder line in the exchange channel, which we show in Fig.\ \ref{fig2}(a). All other diagrams are kinematically suppressed. (We will address the interplay of disorder and Coulomb interaction in more detail in the next Section). Moreover, the fermionic self-energy from the disorder coupling also comes from the exchange channel [Fig.\ \ref{fig2}(b)]. 
For this reason, in this section we only focus on the RG flow of the disorder couplings in the {\it exchange} channel. At the end of this section, we will connect the RG flow of disorder couplings in the exchange channel and Cooper channel via time-reversal symmetry $\mathcal{T}=K$. The RG flow of disorder in the direct channel requires a separate calculation which is not performed here. However, the disorder coupling in the direct channel does not change the momentum or energy of the incoming electron, nor does it renormalize either the fermion Green function or the Coulomb interaction. As such, disorder coupling in the direct channel is not of interest to us in the present work. 

For a Weyl loop system, the replicated action for disorder takes the form
\begin{widetext}
\begin{align}
\mathcal{S}_{\rm dis}=&-\int d\omega d\omega' d^3{{\bf k}} d^3{ {\bf k}'} d^3{{\bf k}''}\frac{W_0({{\bf k},{\bf k}',{\bf k}''})}2[\psi_{i}^\dagger({{\bf k}})\sigma_i^0\psi_{i}({{\bf k}'})]_{\omega}  [\psi_{j}^\dagger({{\bf k}''})\sigma_j^0\psi_{j}({-{\bf k}-{\bf k}'-{\bf k}''})]_{\omega'}\nonumber\\
&-\int d\omega d\omega' d^3{{\bf k}} d^3{ {\bf k}'} d^3{{\bf k}''}\frac{W_x({{\bf k},{\bf k}',{\bf k}''})}2 [\psi_{i}^\dagger({{\bf k}})\sigma_i^x\psi_{i}({{\bf k}'})]_{\omega}  [\psi_{j}^\dagger({{\bf k}''})\sigma_j^x\psi_{j}({-{\bf k}-{\bf k}'-{\bf k}''})]_{\omega'}\nonumber\\
&-\int d\omega d\omega' d^3{{\bf k}} d^3{ {\bf k}'} d^3{{\bf k}''}\frac{W_y({{\bf k},{\bf k}',{\bf k}''})}2 [\psi_{i}^\dagger({{\bf k}})\sigma_i^y\psi_{i}({{\bf k}'})]_{\omega}  [\psi_{j}^\dagger({{\bf k}''})\sigma_j^y\psi_{j}({-{\bf k}-{\bf k}'-{\bf k}''})]_{\omega'}\nonumber\\
&-\int d\omega d\omega' d^3{{\bf k}} d^3{ {\bf k}'} d^3{{\bf k}''}\frac{W_z({{\bf k},{\bf k}',{\bf k}''})}2 [\psi_{i}^\dagger({{\bf k}})\sigma_i^z\psi_{i}({{\bf k}'})]_{\omega}  [\psi_{j}^\dagger({{\bf k}''})\sigma_j^z\psi_{j}({-{\bf k}-{\bf k}'-{\bf k}''})]_{\omega'},
\label{eq22}
\end{align}
\end{widetext}
where $W_{0,x,y,z}$ are four symmetry inequivalent disorder couplings, $i$ and $j$ are replica indices, repeated indices are summed over, and the number of replicas is to be taken to zero at the end of the calculation. By dimension counting, we find that at tree level, the couplings have dimension,
\begin{align}
[W_{a}]=2z-2,~~a=0,x,y,z,
\end{align}
which is again marginal at tree level.

 We have included a momentum dependence into the disorder couplings---even if we start with a constant coupling, a momentum dependence is dynamically generated, as couplings in different channels have different RG flow. {We do assume that the disorder couplings are {\it analytic} functions of momenta. This should generically be the case for short range correlated disorder, such as may be induced by e.g. lattice defects. Assuming analyticity, a Taylor expansion of the function reveals that the leading `constant' piece is the `most relevant' in the renormalization group sense, and we will focus on the renormalization of this piece henceforth.}

{ We also comment briefly on what one might expect in the presence of charged (Coulomb) impurities (donors/acceptors). At first glance, one might think that Coulomb impurities  produce disorder which is long range correlated. However, Coulomb impurities induce electron/hole puddles by self-consistently modifying the local chemical potential, which in turn makes the impurity scattering potential short-ranged via Thomas-Fermi type screening. This is a non-perturbative effect, and has been treated in Ref.\ \onlinecite{skinner} in the context of Dirac semimetals. We expect our present case would follow analogously, although a detailed analysis is beyond the scope of the current work. It was found in Ref.\ \onlinecite{skinner} that the size of the puddle (equivalently, the screening length) $r_s$ is inversely proportional to $\alpha_0 N^{1/3}$, where $\alpha_0$ is the bare Coulomb coupling constant and $N$ is the concentration $N$ of Coulomb impurities.   As long as $r_s$ is smaller than the inverse scale of the nodal ring $k_F^{-1}$, one can treat the scattering potential of the screened Coulomb impurity as momentum-independent and short-range correlated, such that our RG analysis with momentum shell cutoff $\Lambda\ll k_F$ applies. However, if the screening radius is large compared to $k_F^{-1}$, then the momentum dependence of the impurity scattering potential must be taken into account, and a more elaborate analysis using functional RG techniques will be required. This regime is beyond the scope of the present paper, which focuses on short range correlated disorder as explained above. }   
 
  For the disorder coupling exchange channel, we denote 
 \begin{align}
 W_{a}\equiv W_{a}(k_0,k_\theta,k_\theta),
 \end{align}
  where $a=0,x,y,z$.

\begin{figure*}
\includegraphics[width=1.4\columnwidth]{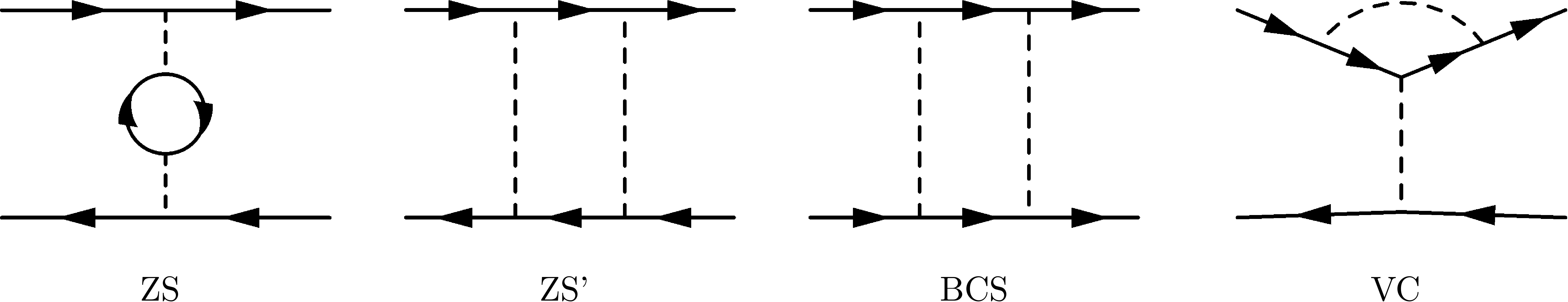}
\caption{Four types of one-loop diagrams that renormalizes the four-fermion coupling. For the disorder problem, the ZS diagram vanishes since it gives a contribution proportional to the number of replicas, which is taken to zero at the end. }
\label{channels_fig}
\end{figure*}
We now analyze the renormalization of the $W$'s. Following the standard terminology of Ref.\ \onlinecite{shankar}, at one-loop level $W$'s receive corrections from ZS, ZS', BCS, and VC type diagrams (shown in Fig.\ \ref{channels_fig}). The contribution from ZS diagram involves a sum over replica indices, and thus vanishes when we take the number of replicas to zero. 
Moreover, one can easily check that only ZS' diagrams benefit from the angular integration; for all other types of diagrams, by momentum conservation in the $xy$ plane the internal momenta cannot be taken at an arbitrary nodal loop angle $\theta$. In the limit $\Lambda\ll k_F$, the contribution from these diagrams can be safely neglected, which greatly simplifies the calculation.

We first consider the ZS' diagram with two $W_0$ lines, which we denote as  $\Gamma^{\rm ZS'}_{00}$ (we follow the same notation for other types of ZS' diagrams):
\begin{widetext}
 \begin{align}
 \Gamma^{\rm ZS'}_{00}=&W_0^2\int\frac{d^3p}{8\pi^3}\frac{\sigma_0^i(i\omega+v_F p_r\sigma_x^i+v_z p_z\sigma_y^i)\sigma_0^i~\sigma_0^j(i\omega+v_F p_r\sigma_x^j+v_z p_z\sigma_y^j)\sigma_0^j}{(\omega^2+v_F^2p_r^2+v_z^2p_z^2)^2}\nonumber\\
 =&\frac{k_F W_0^2\Lambda d\ell}{2\pi^2v_Fv_z}\[{{\sigma_x^i\sigma_x^j}}\int\frac{p_r^2 dp_z}{(p_r^2+p_z^2)^2}+{{\sigma_y^i\sigma_y^j}}\int\frac{p_z^2 dp_z}{(p_r^2+p_z^2)^2}\]=\frac{1}{4}\lambda_0 W_0 d\ell({{\sigma_x^i\sigma_x^j}}+{{\sigma_y^i\sigma_y^j}}),
\end{align}
 where the momentum integral in the first line is for two cylindrical shells sandwiching the line node. Note that for disorder couplings we only integrate over internal momentum but not frequency.  In the spirit of RG, we have set the external frequency $\omega=0$ in the second line. 
 We have introduced dimensionless couplings 
\begin{align}
\lambda_0\equiv k_F W_0/(\pi v_Fv_z).
\label{lambda}
\end{align}
 By a similar procedure, we obtain for all four diagrams of such type,
 \begin{align}
 \Gamma^{\rm ZS'}_{aa}=\frac{1}{4}\lambda_a W_a d\ell({{\sigma_x^i\sigma_x^j}}+{{\sigma_y^i\sigma_y^j}}),~~~a=0,x,y,z.
 \end{align}
 
For $\Gamma^{\rm ZS'}_{0x}$, we have
\begin{align}
 \Gamma^{\rm ZS'}_{0x}= \Gamma^{\rm ZS'}_{x0}=&W_0W_x\int\frac{d^3p}{8\pi^3}\frac{\sigma_0^i(i\omega+v_F p_r\sigma_x^i+v_z p_z\sigma_y^i)\sigma_x^i~\sigma_x^j(i\omega+v_F p_r\sigma_x^j+v_z p_z\sigma_y^j)\sigma_0^j}{(\omega^2+v_F^2p_r^2+v_z^2p_z^2)^2}\nonumber\\
 =&\frac{k_F W_0W_x\Lambda d\ell}{2\pi^2v_Fv_z}\[{\sigma_0^i\sigma_0^j}\int\frac{p_r^2 dp_z}{(p_r^2+p_z^2)^2}+{{\sigma_z^i\sigma_z^j}}\int\frac{p_z^2 dp_z}{(p_r^2+p_z^2)^2}\]=\frac{1}{4}\lambda_0 W_x d\ell({\sigma_0^i\sigma_0^j}+{{\sigma_z^i\sigma_z^j}}).
\end{align}
By the same token, for the rest we have
\begin{align}
 \Gamma^{\rm ZS'}_{0z}= \Gamma^{\rm ZS'}_{z0}=&\frac{1}{4}\lambda_0 W_z d\ell({{\sigma_y^i\sigma_y^j}}+{{\sigma_x^i\sigma_x^j}})\qquad 
 \Gamma^{\rm ZS'}_{0y}= \Gamma^{\rm ZS'}_{y0}=\frac{1}{4}\lambda_0 W_y d\ell({{\sigma_z^i\sigma_z^j}}+{\sigma_0^i\sigma_0^j})\nonumber\\
 \Gamma^{\rm ZS'}_{xz}= \Gamma^{\rm ZS'}_{zx}=&\frac{1}{4}\lambda_x W_z d\ell({{\sigma_z^i\sigma_z^j}}+{\sigma_0^i\sigma_0^j}) \qquad
 \Gamma^{\rm ZS'}_{xy}= \Gamma^{\rm ZS'}_{yx}=\frac{1}{4}\lambda_x W_y d\ell({{\sigma_y^i\sigma_y^j}}+{{\sigma_x^i\sigma_x^j}})\nonumber\\
 \Gamma^{\rm ZS'}_{zy}= \Gamma^{\rm ZS'}_{yz}=&\frac{1}{4}\lambda_z W_y d\ell({\sigma_0^i\sigma_0^j}
+{{\sigma_z^i\sigma_z^j}}).
\end{align}
\end{widetext}
Another potential contribution of disorder comes from the anomalous dimension of the fermion. To see this we compute the fermionic self energy, whose only disorder contribution at one loop [Fig.\ \ref{fig2}(b)] is from the exchange channel, 
\begin{align}
\Sigma(\omega)=&(W_0+W_x+W_z+W_y)\int\frac{d^3p}{8\pi^3}\frac{i\omega}{\omega^2+v_Fp_r^2+v_z^2 p_z^2}\nonumber\\
=&i\omega\times\frac{2\pi k_F}{8\pi^3v_Fv_z}\times \pi\times 2d\ell \times(W_0+W_x+W_z+W_y)\nonumber\\
\equiv& \frac{i\omega}{2}(\lambda_0+\lambda_x+\lambda_z+\lambda_y)d\ell,
\label{222}
\end{align} 
This frequency dependence of self-energy not only renormalizes the dynamical exponent $z$ but also gives rise to an anomalous dimension $\xi$ of the fermion field $\psi$, namely $[\psi_k]=-1-z+\xi$.  From the 
fermion action
\begin{align}
\!\!\!\!\mathcal{S}_f=\int^{\Lambda(\ell)} d\omega dk [\psi^\dagger_k (-i\omega -\Sigma(\omega)+ v_F k_r + v_z k_z)\psi_k
\end{align}
we see that (by enforcing invariance of $v_F$ and the action)
\begin{align}
z=&1
+\frac{1}{2}(\lambda_0+\lambda_x+\lambda_z+\lambda_y)\nonumber\\
\xi=&\frac{1}{4}(\lambda_0+\lambda_x+\lambda_z+\lambda_y).
\label{229}
\end{align}
Taking  the anomalous dimension into account, the dimension of the disorder couplings is
\begin{align}
[W_a]&=[\lambda_a]=2z-2-4\xi=0,
\end{align}
i.e., for the noninteracting problem the disorder coupling (coincidentally) has no anomalous dimension.

We can write down the  the ${\beta}$-functions for the couplings $\lambda$'s as follows:
\begin{align}
\frac{d\lambda_0}{d\ell}=&
\frac{1}{2}(\lambda_0+\lambda_z)(\lambda_x+\lambda_y)\nonumber\\
\frac{d\lambda_x}{d\ell}=&
\frac{1}{4}(\lambda_0+\lambda_z)^2+\frac{1}{4}(\lambda_x+\lambda_y)^2\nonumber\\
\frac{d\lambda_z}{d\ell}=&
\frac{1}{2}(\lambda_0+\lambda_z)(\lambda_x+\lambda_y)\nonumber\\
\frac{d\lambda_y}{d\ell}=&
\frac{1}{4}(\lambda_0+\lambda_z)^2+\frac{1}{4}(\lambda_x+\lambda_y)^2.
\label{26}
\end{align}
Since $\lambda_a \ge 0$, this manifestly flows to strong disorder. We now consider along which trajectory the problem flows to strong disorder, by examining the flow of ratios of the couplings. We have for the ratios
\begin{align}
\frac{d(\lambda_x/\lambda_y)}{d\ell}=&\frac{1}{4\lambda_y}\(1-\frac{\lambda_x}{\lambda_y}\)\[(\lambda_0+\lambda_z)^2+(\lambda_x+\lambda_y)^2\] \nonumber\\
\frac{d(\lambda_0/\lambda_z)}{d\ell}=&\frac{1}{2\lambda_z}\(1-\frac{\lambda_0}{\lambda_z}\)(\lambda_0+\lambda_z)(\lambda_x+\lambda_y) \nonumber\\
\frac{d(\lambda_x/\lambda_0)}{d\ell}=&\frac{1}{4\lambda_0}(\lambda_0+\lambda_z)(\lambda_x+\lambda_y) \nonumber\\
&\times\[\frac{\lambda_0+\lambda_z}{\lambda_x+\lambda_y}+\frac{\lambda_x+\lambda_y}{\lambda_0+\lambda_z}-\frac{2\lambda_x}{\lambda_0}\].
\end{align}
It is easy to check that all ratios flow to $\lambda_0=\lambda_x=\lambda_z=\lambda_y\equiv \lambda$, i.e.\ the flow to strong disorder asymptotes to a fixed trajectory on which all disorder types are equal. Along this fixed trajectory 
\begin{align}
d\lambda/d\ell=2\lambda^2.
\end{align} This is the first main result of our work. Of course, once the problem flows the strong disorder, the weak-coupling RG is no longer valid. 

A comment is in order. So far we have focused on disorder couplings in the exchange channel for a generic $\theta$ (see Fig.\ \ref{fig1}) with no $\theta$ dependence, whose RG flows are driven by ZS' diagrams. However, for a special case $\theta=0$, the diagram is in {\it both} exchange {\it and} direct channel. One can verify that the RG flow for $\lambda$'s at $\theta=0$ thus come from both ZS' diagrams and VC diagrams (which has a positive contribution to its $\beta$-function) and should flow to a larger value. As a result of this, one expect the exchange coupling $W_{a}(\theta)\equiv W_{a}(k_0,k_\theta,k_\theta)$ to have a ``spike" near $\theta=0$ and be flat otherwise. However, it is easy to check that this spike has an angular width in $\theta$ of order of $\sim\Lambda(\ell)$, and thus its feedback to the RG flow of $W_{a}$'s at other (generic) $\theta$'s vanishes as the system flows to low energy.


The renormalization of disorder couplings in the Cooper channel can also be obtained by similar procedures, only the one-loop corrections come from the BCS diagrams ({see Fig.\ \ref{channels_fig}}). The result is
\begin{align}
\frac{d\bar\lambda_0}{d\ell}=&
\frac{1}{2}(\bar\lambda_0+\bar\lambda_z)(\bar\lambda_x-\bar\lambda_y)\nonumber\\
\frac{d\bar\lambda_x}{d\ell}=&
\frac{1}{4}(\bar\lambda_0+\bar\lambda_z)^2+\frac{1}{4}(\bar\lambda_x-\bar\lambda_y)^2\nonumber\\
\frac{d\bar\lambda_z}{d\ell}=&
\frac{1}{2}(\bar\lambda_0+\bar\lambda_z)(\bar\lambda_x-\bar\lambda_y)\nonumber\\
\frac{d\bar\lambda_y}{d\ell}=&
-\frac{1}{4}(\bar\lambda_0+\bar\lambda_z)^2-\frac{1}{4}(\bar\lambda_x-\bar\lambda_y)^2,
\end{align}
where $\bar\lambda$'s are dimensionless couplings in the Cooper channel. Note that the only difference from (\ref{26}) is an additional minus sign for $\bar\lambda_y$. This is a result of time-reversal symmetry $\mathcal{T}=K$ -- Cooper channel and exchange channel are related by an time-reversal operation on {\it one} of the fermion lines, and the relative minus sign between $\lambda_y$ and $\bar\lambda_y$, which both couple via $\sigma^y$, is due to the fact that under TR, $\sigma^y\to -\sigma^y$. Therefore under RG flow the TR symmetry is indeed preserved.

We finally remark that the RG flow in the direct channel can also be computed, and at one-loop level the correction comes from VC diagrams. However, as we argued, direct channel scattering affects neither the fermion Green function nor the Coulomb interaction, and so is not of interest to us here. 

\section{Feedback effects  between Coulomb interaction and disorder}
\label{sec:4}

In this section we combine effects of disorder and Coulomb interaction.~\cite{jinwuye, herbut2001, fg1988} First, including both effects, the dynamical exponent $z$ and anomalous dimension $\xi$ are modified to, according to Eqs.\ (\ref{117},~\ref{229}):
\begin{align}
z=&1-\frac\alpha{8\pi^2}F_1(a\eta)
+\frac{1}{2}(\lambda_0+\lambda_x+\lambda_z+\lambda_y)\nonumber\\
\xi=&\frac{1}{4}(\lambda_0+\lambda_x+\lambda_z+\lambda_y).
\label{dex}
\end{align}
Thus, the couplings have scaling dimensions
\begin{align}
[W_a]=&2z-2-4\xi=-\frac{\alpha}{4\pi^2} F_1(a\eta),\nonumber\\
[\lambda_a]=&[W_a]-[\eta]=-\frac{\alpha}{4\pi^2} F_{12}(a\eta),\nonumber\\
[\bar\alpha]=&z-4\xi\nonumber\\
=&1-\frac\alpha{8\pi^2}F_1(a\eta)-\frac{1}{2}(\lambda_0+\lambda_x+\lambda_z+\lambda_y),
\label{400}
\end{align}
where the second line follows from the definition \eqref{lambda} and the last line of \eqref{119}, and here $F_{12}\equiv (F_1+F_2)/2$.

{We now investigate the direct interplay of disorder and interactions through diagrams involving `mixed' disorder and interaction lines}. At one-loop level,  the relevant diagrams are of the type of those from Fig.\ \ref{channels_fig}, only with one dashed line (disorder) replaced with a wavy line (Coulomb interaction). For our purposes, we are only concerned with the long range Coulomb interaction (corresponding to small momentum transfer) and exchange-channel disorder. 

{The calculations are greatly simplified because of kinematic constraints, which cause most of the diagrams in question to make vanishing contribution. First, consider the ZS diagram with one wavy line and one dashed line. For external in-plane momenta at $k_0$ and $k_\theta$, the {\it internal} in-plane momenta of the fermions in the bubble are restricted to be at $k_0$ and $k_\theta$ as well. Hence, kinematic constraints cause the ZS diagram to make a vanishing contribution, since there are no internal momenta to integrate over. Now consider the ZS' and BCS diagrams with one wavy line and one dashed line. The wavy line (Coulomb) propagator is peaked at  momentum transfer zero. Thus, the momentum integration in these diagrams predominantly comes from internal momenta approximately equal to the external momenta. There diagrams therefore are again kinematically suppressed (similar to the discussion in Ref.\ \onlinecite{NandkishorePara}), and make vanishing contribution. The VC diagram with a Coulomb line dressing a disorder vertex makes vanishing contribution for analogous reasons. The only one-loop diagram that is not suppressed by kinematic constraints is the VC diagram where a disorder line dresses a Coulomb vertex. In this diagram, the {\it internal} momenta are unrestricted, and integration over internal momenta generates a log divergence suitable for resummation in an RG procedure. A consideration of the replica structure of this diagram reveals that it corresponds to a  renormalization to the long-range Coulomb interaction. We therefore conclude that at one loop level, diagrams involving mixed disorder and Coulomb lines produce a log divergent correction to the Coulomb interaction (and hence contribute to the Coulomb $\beta$ function), but do not make any log divergent contributions to the disorder couplings. That is, at one loop level, diagrams with mixed disorder and interactions lines do not contribute to the $\beta$ functions for the $\lambda$s, which from Eqs.\ (\ref{26},\ref{400}) then simply take the form}
\begin{align}
\frac{d\lambda_0}{d\ell}=&\lambda_0\[-\frac{\alpha}{4\pi^2}F_{12}(a\eta)\]+
\frac{1}{2}(\lambda_0+\lambda_z)(\lambda_x+\lambda_y)\nonumber\\
\frac{d\lambda_x}{d\ell}=&\lambda_x\[-\frac{\alpha}{4\pi^2}F_{12}(a\eta)\]+
\frac{1}{4}(\lambda_0+\lambda_z)^2+\frac{1}{4}(\lambda_x+\lambda_y)^2\nonumber\\
\frac{d\lambda_z}{d\ell}=&\lambda_z\[-\frac{\alpha}{4\pi^2}F_{12}(a\eta)\]+
\frac{1}{2}(\lambda_0+\lambda_z)(\lambda_x+\lambda_y)\nonumber\\
\frac{d\lambda_y}{d\ell}=&\lambda_y\[-\frac{\alpha}{4\pi^2}F_{12}(a\eta)\]+
\frac{1}{4}(\lambda_0+\lambda_z)^2+\frac{1}{4}(\lambda_x+\lambda_y)^2,
\label{226}
\end{align}

{Meanwhile, the log divergent diagram involving a disorder line $\lambda_0$ dressing the Coulomb vertex produces a vertex correction}
\begin{align}
&\Gamma_{c0}^{\rm VC}=\frac{-e^2}{aq_r^2+\frac{1}{a}q_z^2}\times2W_0\int \frac{d^3p}{8\pi^3}\nonumber\\
&\times\frac{\sigma^i_0\sigma^j_0(i\omega+v_F p_r \sigma_x^i+v_z p_z \sigma_y^i)\sigma^i_0(i\omega+v_F p_r \sigma_x^j+v_z p_z \sigma_y^j)\sigma_0^i}{(\omega^2+v_Fp_r^2+v_z^2p_z^2)^2}\nonumber\\
&=\frac{-e^2}{aq_r^2+\frac{1}{a}q_z^2}\times2W_0\int \frac{d^3p}{8\pi^3}\frac{-\omega^2+v_F^2p_r^2+v_z^2 p_z^2}{(\omega^2+v_Fp_r^2+v_z^2p_z^2)^2}\nonumber\\
&=\frac{-e^2}{aq_r^2+\frac{1}{a}q_z^2}\lambda_0d\ell,
\end{align}
Via similar calculations, we find that
\begin{align}
&\Gamma_{c0}^{\rm VC}+\Gamma_{cx}^{\rm VC}+\Gamma_{cz}^{\rm VC}+\Gamma_{cy}^{\rm VC}\nonumber\\
&=\frac{-e^2}{aq_r^2+\frac{1}{a}q_z^2}(\lambda_0+\lambda_x+\lambda_z+\lambda_y)d\ell,
\label{422}
\end{align}
and  the RG equation for ${\bar \alpha}$ is, 
\begin{align}
\frac{d{\bar \alpha}}{d\ell}=&{\bar \alpha}\[1-\frac{\alpha}{8\pi^2}F_1(a\eta)-\frac{1}{2}(\lambda_0+\lambda_x+\lambda_z+\lambda_y)\]\nonumber\\
&+{\bar \alpha}\[-\frac{1}{2}{\bar \alpha}\(\frac{1}{2a\eta}+2a\eta\)+(\lambda_0+\lambda_x+\lambda_z+\lambda_y)\]\nonumber\\
=&{\bar \alpha}\[1-\frac{1}{2}{\bar \alpha}\(\frac{1}{2a\eta}+2a\eta\)-\frac{\alpha}{8\pi^2}F_1(a\eta)\right.\nonumber\\
&\left.~~~~+\frac12(\lambda_0+\lambda_x+\lambda_z+\lambda_y)\].\label{35}
\end{align}
{Note that in our analysis of the clean problem we ignored the potential renormalization of the dynamics of the boson $\phi$ due to the absence of a logarithmic contribution from the polarization bubble~\cite{HuhMoonKim}. In principle one may wonder if this is still a safe assumption in the presence of disorder. However, our consideration of all diagrams at one loop has revealed that the only effect of disorder on Coulomb interaction is the renormalization of $\bar\alpha$. Therefore it is safe to continue to neglect the dynamics of $\phi$ even in the disordered case, at least to one loop.}

The RG equation for the anisotropy ratio $a\eta$ is the same as in the clean case, since as we have seen, disorder does not modify $a$ or $\eta$.
\begin{align}
\frac{d(a\eta)}{d\ell}&=a\eta\[\frac{1}{2}{\bar \alpha}\(\frac{1}{2a\eta}-2a\eta\)+\frac{\alpha}{8\pi^2}\(F_2(a\eta)-F_1(a\eta)\)\].
\label{36}
\end{align}

Let's analyze this set of RG equations in (\ref{226},~\ref{35},~\ref{36}). We assume and then justify we can safely neglect $\sim\alpha$ terms in Eq.\ \eqref{226}. Then the RG flow of $\lambda$'s is unaffected by Coulomb interaction.
Since $\lambda_{0,x,y,z}=\lambda$ all flow to strong coupling, from Eq.\ \eqref{35} ${\bar \alpha}$ also flows to strong coupling. However, the ratio of ${\bar \alpha}/\lambda$ flows to zero. This can be seen from the $\beta$-function for ${\bar \alpha}/\lambda$. We use 
Eqs.\ \eqref{226} and \eqref{35} and $\lambda_{0,x,y,z}=\lambda$ to obtain,
\begin{align}
\frac{d({\bar \alpha}/\lambda)}{d\ell}=&\frac{1}{\lambda}\frac{d\bar\alpha}{d\ell}-\frac{\bar\alpha}{\lambda^2}\frac{d\lambda}{d\ell}\nonumber\\
=&\frac{\bar \alpha}{\lambda}\[1-\frac{\bar \alpha}2\(\frac{1}{2a\eta}+2a\eta\) - \frac{\alpha}{8\pi^2}F_1(a\eta)\].
\end{align}
As ${\bar \alpha}$ flows to strong coupling, the ratio ${\bar \alpha}/\lambda$ flows to zero. 
The energy scale $\Lambda(\ell^*)$ at which $\lambda$ and $\bar\alpha$ diverges is given by integrating $d\ell/d\lambda=1/(2\lambda^2)$, and doing so we find $\ell^*={1}/{(2\lambda_b)}$, where $\lambda_b$ is the (approximate) bare value of the disorder couplings $\lambda_{0,x,y,z}$. We thus find 
 \begin{align}
 \Lambda(\ell^*)=\Lambda_0 e^{-1/(2\lambda_b)},
 \label{355}
 \end{align}
 where $\Lambda_0$ is the bare cutoff of the theory that is of the order of band width.
 From this we know that the Coulomb coupling $\alpha\sim\bar\alpha \Lambda(\ell)$ flows to strong coupling, but is much weaker than the disorder strength.  This is the second main conclusion of our work.
 
 As for the anisotropy parameter $a\eta$, we see from Eq.\ \eqref{36} that the fixed point remains close to $1/2$, and is corrected by the $\sim\alpha$ term. The fixed point is given by, roughly
 \begin{align}
 a\eta=\frac12+ O\[\exp{\(-\frac1{2\lambda_b}\)}\],
 \end{align}
 where we have used the fact that $F_{1,2}(1/2)\approx-3.7=O(1)$.

\section{Weyl loop vs. Dirac loop}
\label{sec:5}

So far we have considered a ``minimal model" of the nodal-line semimetal in which the nodal line is two-fold degenerate, i.e., a Weyl loop. 
It is also interesting to consider a four-fold degenerate nodal line, i.e., a Dirac loop system,
\begin{align}
\mathcal{H}_0=\frac{k_x^2+k_y^2-k_F^2}{2m}\Gamma_1+v_z k_z \Gamma_2,
\label{dloop}
\end{align}
where $\{ \Gamma_a,\Gamma_b\}=2\delta_{ab}$, where the $\Gamma$ are four dimensional matrices. Experimentally, Ca$_3$P$_2$, CaAgP, and CaAgAs~\cite{NeupertHasan,Takenaka, Loop1} are found to realize the Dirac loop band structure.

The four bands crossing at the Dirac loop can be thought of as coming from two orbital times two spin degrees of freedom. With negligible spin-orbit interaction (for example in Ca$_3$P$_2$ and CaAgP), each orbital band is spin degenerate, and the Dirac loop is simply two copies of Weyl loops. Further, if only nonmagnetic disorder is included, the problem with disorder and Coulomb interaction essentially reduces to the one we considered above for the Weyl loop aside from an unimportant factor of $2$ in the term representing screening of the Coulomb interaction, which does not change any of our results. 

The situation becomes more involved if spin-orbit interaction cannot be neglected, for example in CaAgAs. In this case more types of disorder need to be included. Generically, there exist 16 linearly independent rank-4 matrices: an identity matrix, five $\Gamma_a$'s, where $a=0,1,2,3,5$, and ten $\Gamma_{ab}$'s where $\Gamma_{ab}\equiv \frac{1}{2i}[\Gamma_a,\Gamma_b]$. Each independent $\Gamma$ matrix corresponds to a disorder type and generically all 16 of $\lambda$'s needs to be treated on an equal footing. Analogous to Eq.\ \eqref{26}, one can verify that the coefficients of the $O(\lambda^2)$ terms in the $\beta$-functions are all positive. To see this, recall that in the ZS' diagrams the matrix structure follows the basic form of $(\Gamma_\alpha^i\Gamma_\beta^i\Gamma_\gamma^i)(\Gamma_\gamma^j\Gamma_\beta^j\Gamma_\alpha^j)$, where $\Gamma_\alpha = \mathbb{I}, \Gamma_a$ or $\Gamma_{ab}$, and the remaining momentum integral is positive. From the Hermiticity of $\Gamma_{\alpha}$, the above matrix products can be written as $(\Gamma_\alpha^i\Gamma_\beta^i\Gamma_\gamma^i)(\Gamma_\alpha^j\Gamma_\beta^j\Gamma_\gamma^j)^\dagger=+\Gamma_\delta^i \Gamma_\delta^j$, where $\delta$ corresponds to a certain type of disorder. Therefore, we expect that all 16 types of disorder flow to strong coupling under RG, and due to the feedback effects,  Coulomb interaction is driven toward strong coupling but remains asymptotically weaker. 
However, the ratios of different disorder coupling generally do not approach one, and require a direct solution of 16 coupled equations, which we leave to future work.

\section{Discussion and conclusion}
\label{sec:6}
In this work, we analyzed the interplay of Coulomb interactions and disorder in the nodal-line semimetals using a weak coupling RG approach. As we showed, the combination of the vanishing density of states and the Fermi-surface-like kinematics makes the nodal-line band structure an interesting playground for RG studies. At  tree level, both Coulomb interaction and disorder coupling are marginal. We showed that for a Weyl loop material, all types of disorder  flow to strong coupling, and asymptote to a fixed trajectory along which all disorder strengths are equal. The feedback effect from disorder in turn drives the Coulomb interaction to strong coupling. This is in contrast to the clean case, in which the Coulomb interaction vanishes under RG~\cite{HuhMoonKim}. However, the Coulomb interaction (in the disordered case, when it flows to strong coupling) grows asymptotically more slowly than disorder, such that the ratio of Coulomb interactions to disorder flows to zero. Extrapolating our results to strong coupling, it seems likely that the strong coupling low-energy theory is described by a non-interacting nonlinear sigma model. Of course, this conclusion is based on an extrapolation to strong coupling, where the weak coupling RG is ill controlled, and a rigorous treatment of the strong coupling regime is left to future work. Our results should apply unchanged to Dirac loop materials with negligible spin orbit coupling and only non-magnetic disorder. For Dirac loop materials with appreciable spin orbit coupling, or with magnetic disorder, qualitatively similar results should obtain, but the different disorder strengths need not become equal as the problem flows to strong coupling. 

Our results should be directly testable within ARPES experiments on Weyl or Dirac loops measuring the energy dependence of the quasiparticle dynamic exponent, residue, and lifetime. We collect our predictions below
From the calculations above: the dynamical exponent $z$ is asymptotically given by
\begin{align}
z=1+2\lambda.
\end{align}
Instead of introducing an anomalous dimension to the fermion field $\psi$, we can also keep $[\psi]$ as its engineering dimension and replace the fermion action by $\mathcal{S}_f=\int d\omega d^3 k \psi^\dagger Z^{-1}(-i\omega+\epsilon_{\bf k})\psi$. This way the anticommutation relation between creation and annihilation operators are preserved along the RG flow, and $Z$ is nothing but the quasiparticle residue.
It is easy to see that the dimension of $Z$ is proportional to the anomalous dimension and, asymptotically,
\begin{align}
\frac {dZ}{d\ell}=-2\xi Z=-2\lambda Z,
\end{align}
where the flow of $\lambda$ is given by $d\ell/d\lambda=1/(2\lambda^2)$. This tells us how $Z$ vanishes in the low energy limit. We caution that the predictions for renormalization of dynamic exponent and quasiparticle reside are only valid in the perturbative regime, when $\lambda < 1$. Finally, the growth of disorder should give rise to a non-zero quasiparticle lifetime, which may be  estimated by the energy scale at which disorder becomes strong. According to Eq.\ \eqref{355}, 
\begin{align}
\tau\sim e^{1/(2\lambda_b)}/\Lambda_0 ,
\end{align}
where $\lambda_b$ is the rough bare strength of the disorder and $\Lambda_0$ is the bare cutoff of the theory which is given by the energy scale at which the bandstructure departs appreciably from Eq.\ \eqref{wloop}.

Another measurable quantity is the electronic compressibility $K(\mu)=-\partial^2F/\partial \mu^2$, where $F=-(T/V)\log \mathcal{Z}$. In the clean and free case, the Dirac points along the nodal loop give rise to a linear density of states, and thus in the presence of a small doping $\mu$ (which also gives a natural IR scale), the compressibility is $K_0(\mu)\sim \mu$. This can also be obtained by a scaling argument by noting that $[F]=z+2$, $[K]=z$, and $z=1$ in the free fermion case.
%
 Calculating the free energy within perturbation theory in weak $\alpha$ and $\lambda$, we find that to leading order the fractional change in the compressibility is given by,
\begin{align}
\frac{\delta K(\mu)} {K_0(\mu)}=-\frac{\alpha F_{1}}{8\pi^2}\log\frac{\Lambda}{\mu},
\end{align}
i.e., at leading order, disorder does not directly renormalize compressibility (for details see the Appendix). 
Of course, once the system flows to strong disorder the perturbative calculation ceases to apply, and other effects (such as generation of low energy density of states by disorder~\cite{dassarma}) may come to dominate. 

Finally, an accurate description of transport experiments (for results with weak disorder, see a recent work Ref.\ \onlinecite{Carbotte2017}) at low temperatures will require a sigma model, and lies beyond the scope of the present work. The specific non-linear sigma model expected to govern the strong coupling physics depends on the symmetries of the model in question. For the Weyl loops discussed here (or for Dirac loops with negligble spin orbit coupling and only non-magnetic disorder), the sigma model would be in the orthogonal class. This class admits of a localized and a metallic phase, with an Anderson localization transition that can be described using standard techniques, see e.g. Ref.\ \onlinecite{EversMirlin}. Given that the bare Hamiltonian contains a long range Coulomb interaction, Weyl loops may thus provide a rare example of an analytically tractable many body localization transition {driven by tuning disorder strength}. {Of course, while most works on many body localization focus on systems at high temperature, this would be a zero temperature transition. Also, irrelevance of the Coulomb interaction at the sigma model critical point would have to be demonstrated, likely requiring a Finkelstein-type analysis \cite{finkelstein}, but our results indicating that the Coulomb interaction becomes asymptotically weaker than disorder as the problem flows to strong coupling are encouraging in this regard. }
Meanwhile for strongly spin orbit coupled systems where the time reversal symmetry operator squares to minus one, the sigma model would be in the symplectic class, and the discussion follows exactly the corresponding discussion in Ref.\  \onlinecite{NandkishorePara}. 

Our analysis of disorder has thus far ignored rare region effects. Rare region effects are known to dominate the low energy physics of disordered Weyl semimetals \cite{NandkishoreHuseSondhi, pixleyhuse, pixleychou, OstrovskyWeyl, GurarieInstanton}. However, in that case the disorder is perturbatively irrelevant \cite{Fradkin, Fradkin2, SyzranovARCMP}, so the only effect of disorder comes through rare regions. In the present scenario, disorder is perturbatively {\it relevant} (if only marginally so), and so we expect rare region effects to be less important. A detailed analysis of rare regions is left to future work. 

Finally, we note that while a localized phase can obtain for strong disorder at {\it zero temperature},  it is generally believed that at finite temperature the Coulomb interaction is `too long range' to allow many body localization \cite{Burin, yaodipoles}. This understanding has recently been called into question by Ref.\ \onlinecite{NandkishoreSondhiForthcoming}, which provided explicit examples of many body localized phases at finite temperature with Coulomb interactions, at least up to rare regions. Understanding whether the localized phase discussed herein survives at non-zero temperatures, and if not, how the conductivity and characteristic length scales scale with temperature, would also be an interesting topic for future work. 

\acknowledgements We thank S. Sur for useful comments on the manuscript.  RN would like to thank S.A. Parameswaran for a collaboration on related work. YW acknowledges useful discussions with E. Fradkin and S. Ryu. This work was supported by the Gordon and Betty Moore Foundation's EPiQS Initiative through Grant No.\ GBMF4305 at the University of Illinois (Y.W.), and by the Sloan Foundation through a Sloan Research Fellowship (RN). 

\appendix

\section{Effect of weak disorder on compressibility}

\begin{figure}
\includegraphics[width=0.7\columnwidth]{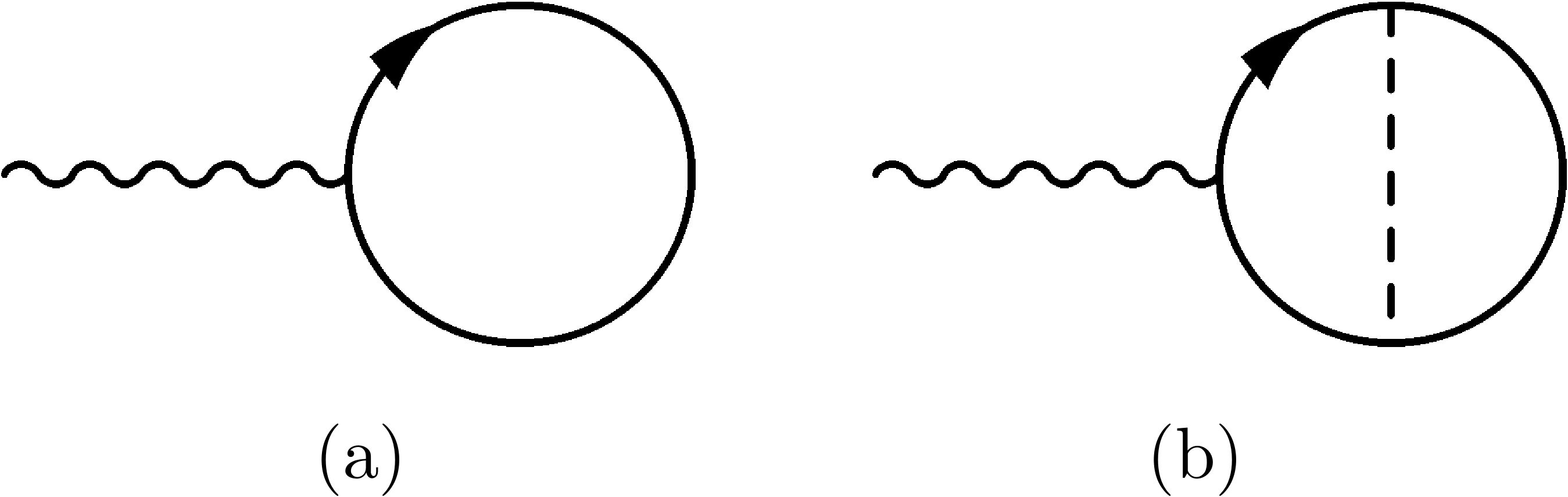}
\caption{The tadpole diagrams for the total particle number (a) for the clean case, and (b) for the disordered case.}
\label{tadpole}
\end{figure}

In this appendix we perform a perturbative calculation of the disorder effect on compressibility for an non-interacting doped Weyl loop semimetal. We show that weak disorder alone does not introduce a change in  compressibility on the leading perturbative level. Note that for a small doping $\mu$, disorder ultimately flows to strong coupling 
and may change the density of states and compressibility. However, these effects cannot be captured within perturbation theory.

The compressibility is given by 
\begin{align}
K(\mu)=\frac{\partial n(\mu)}{\partial{\mu}},
\end{align}
and in diagrammatic language the particle density $n$ is given by the tadpole diagram, shown in Fig.\ \ref{tadpole}. Below we compute the tadpole diagram for the clean case, and then its first-order correction by disorder.

For the clean case, the tadpole diagram Fig.\ \ref{tadpole}(a) is given by 
\begin{align}
n_0(\mu)=&\int \frac{k_F d^2 k}{(2\pi)^2} n(\bm k)
\end{align}
where
\begin{align}
n(\bm k)=&\int\frac{d\omega}{2\pi} \Tr\[\frac{e^{i\omega 0_+}(i\omega-\mu + \bm k\cdot \bm \sigma)}{(\omega-i\mu)^2+k^2} \]\nonumber\\
=&\int\frac{d\omega}{\pi} \frac{e^{i\omega 0_+}(i\omega-\mu)}{(\omega-i\mu)^2+k^2},
\label{s3}
\end{align}
where the additional factor $e^{i\omega 0_+}$ ensures the time ordered operator is also normal ordered. This additional factor enable us to close the contour from the upper half plane. For convenience we have set $v_F=v_z=1$.

We perform the integral in \eqref{s3} by residue theorem. For definiteness we set $\mu<0$ (the case with $\mu>0$ can be inferred from particle-hole symmetry). The poles of the integrand are $\omega_{1,2}=-i|\mu|\pm i k$, ($k=|\bm k|>0$). In this case, there exists a pole $\omega_1$ on the upper half-plane only if $k>|\mu|$. Thus,
\begin{align}
n(\bm k)=\begin{cases}
1,& k>|\mu| \\
0,& k\leq|\mu|,
\end{cases}
\end{align}
consistent with the picture of a Dirac fermion with its lower band partially filled. It is straightforward to compute its contribution to the compressibility, which is nothing but the density of states of free Dirac fermions.

Now we move to the diagram in Fig.\ \ref{tadpole}(b). We have
\begin{align}
\delta n(\mu)=& \tilde\lambda \Tr\[ \int\frac{d\omega d^2k}{(2\pi)^3} e^{i\omega 0_+}\frac{i\omega + \mu +\bm k\cdot \bm \sigma}{(\omega-i\mu)^2+k^2}\right.\nonumber\\
&\times\left. \int \frac{d^2 q}{(2\pi)^2} \frac{i\omega + \mu +\bm q\cdot \bm \sigma}{(\omega-i\mu)^2+q^2}\times \frac{i\omega + \mu +\bm k\cdot \bm \sigma}{(\omega-i\mu)^2+k^2}\]\nonumber\\
=& -2i\tilde\lambda  \int\frac{d\omega d^2k d^2q}{(2\pi)^5} e^{i\omega 0_+}\nonumber\\
&\times\frac{(\omega -i \mu)\[(\omega-i\mu)^2-k^2\]}{[(\omega-i\mu)^2+k^2]^2[(\omega-i\mu)^2+q^2]},
\end{align}
where $\tilde\lambda$ is a prefactor proportional to the disorder strength. We still set $\mu<0$ for the ease of completing the contour in the upper half plane, and the poles on the upper half plane are $\omega_1=-i|\mu|+ik$ (for $k>|\mu|$, double pole), and $\omega_2=-i|\mu|+iq$ (for $q>|\mu|$). Using the residue theorem, we find
\begin{widetext}
\begin{align}
\delta n\propto& \int_{q>|\mu|} d^2q d^2k \frac{-(k^2+q^2)}{2(k^2-q^2)^2}+ \int_{k>|\mu|} d^2q d^2k \frac{\partial}{\partial \tilde\omega}\[\frac{\tilde\omega (\tilde\omega^2-k^2)}{(\tilde\omega+ik)^2(\tilde\omega^2+q^2)}\]\bigg|_{\tilde\omega=ik} \nonumber\\
=&  \int_{q>|\mu|} d^2q d^2k \frac{-(k^2+q^2)}{2(k^2-q^2)^2} +  \int_{k>|\mu|} d^2q d^2k \frac{(k^2+q^2)}{2(k^2-q^2)^2}  \nonumber\\
=& 0\ !
\end{align}
\end{widetext}
Thus, $\delta K(\mu) = \partial \delta n(\mu) / \partial \mu =0$, i.e., at leading order weak disorder does not contribute to the particle number or compressibility.

\bibliography{CAREER_bib}

\end{document}